\documentclass[useAMS,usenatbib]{mn2e}

\usepackage[T1]{fontenc}
\usepackage[latin9]{inputenc}
\usepackage{float}
\usepackage{graphicx}
\usepackage{amssymb}
\usepackage{esint}

\makeatletter

\DeclareRobustCommand{\greektext}{%
  \fontencoding{LGR}\selectfont\def\encodingdefault{LGR}}
\DeclareRobustCommand{\textgreek}[1]{\leavevmode{\greektext #1}}
\DeclareFontEncoding{LGR}{}{}
\DeclareTextSymbol{\~}{LGR}{126}

\begin{document}

\title{Dark Energy as Double N-flation\textemdash{}Observational Predictions}

\author[J. Richard Gott, III and Zachary Slepian]{J.Richard Gott, III \thanks{E-mail: jrg@astro.princeton.edu
    (RG)}, and Zachary Slepian\thanks{E-mail: zslepian@Princeton.edu}\\ 
Department of Astrophysical Sciences, Princeton University, Princeton, NJ, 08544\\
}
\maketitle
Submitted: 7 December 2011

\begin{abstract}
We propose a simple model for dark energy useful for comparison with
observations. It is based on the idea that dark energy and inflation
should be caused by the same physical process. As motivation, we note
that Linde\textquoteright{}s simple chaotic inflation $V=\frac{1}{2}m^{2}\phi^{2}$
produces values of $n_{s}=0.967$ and $r=0.13$, which are consistent
with the WMAP 1\textgreek{sv} error bars. We therefore propose $V=\frac{1}{2}m_{2}^{2}\phi_{2}^{2}+\frac{1}{2}m_{1}^{2}\phi_{1}^{2}$
with $m_{1}\sim10^{-5}$ and $m_{2}\leq10^{-60}$, where $c=1=\hbar$
and the reduced Planck mass is set to unity. The field $\phi_{1}$
drives inflation and has damped by now ($\phi_{1,0}=0$), while $\phi_{2}$
is currently rolling down its potential to produce dark energy. Using
this model, we derive the formula $\delta w(z)\equiv w(z)+1=\delta w_{0}(H_0/H(z))^2$
via the slow-roll approximation. Our numerical results from exact
and self-consistent solution of the equations of motion for $\phi_2$ and the Friedmann equations support this
formula, and it should hold for any slow-roll dark energy.

Our potential can be easily realized in N-flation models with many
fields, and is easily falsifiable by upcoming experiments\textemdash{}for
example, if Linde\textquoteright{}s chaotic inflation is ruled out.
But if $r$ values consistent with Linde\textquoteright{}s chaotic
inflation are detected then one should take this model seriously indeed. 
\end{abstract}

\begin{keywords}
cosmological parameters -- cosmology: theory, cosmology: dark energy, equation of state, inflation, early Universe observations
\end{keywords}

\section{Introduction}

Measuring dark energy is one of the most exciting problems in cosmology
today. There are a number of expensive programs underway to measure
$w,$ the ratio of the pressure to the energy density in dark energy.
If dark energy is a pure cosmological constant, then $w=-1$ for all time.
Currently, measurements of $w$ are compared to a toy model in which
$w$ changes linearly with expansion factor $a$: $w=w_{0}+w_{a}(1-a)$. Observational programs are judged by a figure of merit which includes
their ability to measure the quantities $w_{0}$ and $w_{a}$ in this
toy model, also known as the Chevallier-Polarski-Linder (CPL) parametrization (Albrecht et al. 2006, Chevallier \& Polarski 2001, Linder
2003). 

If one parametrizes $w(a)$ in terms of the above toy model,
the current $1\sigma$ limits from the 7-year WMAP data combined with
BAO+$H_0$+SN are $w_{0}=-.93\pm.13,$ and $w_{a}=-.41\pm.72$ (Komatsu
et al. 2010). These values are consistent at the $1\sigma$ level
with $w_{0}=-1$ and $w_{a}=0,$ which would be a pure cosmological
constant. 
\newpage
In the near future, the Sloan III survey should measure the Baryon Acoustic
Oscillation (BAO) scale to high accuracy using LRG's (Luminous Red Galaxies), and should
lead to a measurement of $w_{0}$ to an accuracy of $3\%$ (SDSS-III 2008). Using
this same dataset, we can use genus topology to measure $w_{0}$ independently
to an accuracy of $5\%$ (Zunckel, Gott, \& Lunnan 2010, Park \&
Kim 2009). Supernova studies can achieve similar results (c.f. Riess
et al. 1998, Riess et al. 2007, Albrecht et al. 2006). 

In the longer term, the $Euclid$
and $WFIRST$ (Wide-Field Infrared Survey Telescope) satellite missions may be able to achieve an accuracy of
$1\%$ (Cimatti et al. 2009, Blandford et al 2010). Such observations may continue to point to $w_{0}=-1$ and
$w_{a}=0$ with higher and higher accuracy, which would be an important
result, but would leave us still in the dark as to the exact nature
of dark energy. 

More exciting would be if a detectable difference $\delta w_{0}$
between $w_{0}$ and $-1$ is found (i.e. $\delta w_{0}=w_{0}+1$).
For this reason, it is desirable to consider simple models, consistent
with current data and falsifiable in principle in the near future, in
which there is a chance that such a detectable difference may be found.
Such models may be used as a guide for interpreting the observations.
We therefore propose such a simple model, based on the idea that inflation
and dark energy come from the same physical mechanism. This hypothesis
is not present in most of the standard theories popular today. 

Today, the most popular theory for dark energy is that a string landscape
exists with many metastable vacua with different values of $V_{0}$.
In this picture, we are currently sitting at the bottom of a potential
well whose low point has a vacuum energy density of $V_{0}$. Thus,
the accelerated expansion we see in the Universe today is attributed
to our current static location at a metastable well in the potential,
while in contrast, we have evidence that the accelerated expansion
we see in the early Universe is due to slow-roll inflation, where
the field is slowly rolling down a potential hill.

Furthermore, for dark energy today, there is supposed to be a complicated
potential that is a function of many fields $\phi_{i}$ rather than
merely the one field in slow-roll inflation. The kinetic energy in
these $\phi_{i}$ fields has damped out and they are no longer changing.
A problematic feature of this model is the value of $V_{0}$: $10^{-120}$
(with $c=1=8\pi G=\hbar),$ an extraordinarily small number. The common
solution to this problem is to propose that there are of order $10^{500}$
vacuum states with values of $-1<V_{0}<1.$ Then one invokes anthropic
effects to argue that it would be difficult for intelligent life to
evolve unless $V_{0}\sim10^{-120}$ (cf. Weinberg 1987, Vilenkin 2003).
This is not as satisfying to some physicists as an actual prediction
of the amount of dark energy we observe made directly from the physical
model. 

A final problem with this model is the appearance of Boltzmann brains
(see discussion in Gott 2008 and references therein). Briefly, in
the far future, the Universe comes to a finite Gibbons and Hawking
temperature $T=1/(2\pi r_{0})$, where $r_{0}=(3/V_{0})^{1/2}$, and
Boltzmann brains appear. While this problem may be manageable depending
on what measure one uses (c.f. deSimone et al. 2010), it is still
a problem that must be addressed in the popular $V_{0}$ model. 

This paper will be structured as follows. In section 2, we discuss
previous models of dark energy and attempts to unify dark energy with
inflation, as well as observational evidence for inflation. In section
3, we discuss Linde's chaotic inflation. In section 4, we propose
our double inflation potential based on Linde's model. In sections
5 and 6, we show how our potential might be realized as an effective
N-flationary potential arising from many axion fields, and explain
why this may be regarded as preferable. In section 7, we estimate the probability of  different deviations from $w=-1$.  In section 8, we describe
the numerical approach we used to obtain exact values of $\delta w_{0}$
and $w(z)$, and in section 9 we present  the 
results we obtain.  In section 10, we conclude.

We use $c=1=\hbar=8\pi G$
throughout, where $G$ is Newton's constant. In these units, the reduced
Planck mass $m_{pl}$ $=(\hbar c/8\pi G)^{1/2}=2.44\times10^{18}\ GeV$
is equal to unity. These units follow Linde (2002).

\section{Previous Dark Energy and Unification Models}
\subsection{Dark Energy}

While a pure cosmological constant $w\equiv-1$ is the simplest model
that explains current observations (see Li, Li \& Zhang 2010), there
are numerous other dark energy models, which we briefly discuss here
to place our own approach in context. Broadly, there are scalar-field
models such as quintessence, k-essence, a tachyon field, phantom dark
energy, dilatonic dark energy, and a Chaplygin gas. Alternatively,
there are also proposals that the Universe's accelerating expansion
may stem from modified gravity, effects of physics above the Planck
energy, or backreactions of cosmological perturbations. For these
models, we direct the reader to the review of Copeland et al.
(2006); here we will focus on summarizing the scalar-field models, since our model falls into this category.

Quintessence is characterized by a scalar field that rolls down its potential to produce the accelerated expansion we observe. Slow-roll solutions are
possible if certain conditions are satisfied (see Copeland et al. 2006), and both exponential and power-law type potentials
are popular. One possible difficulty with quintessence models is that
if the field couples to ordinary matter this could lead to time dependence
of the constants of nature. In contrast to quintessence,
for k-essence, the expansion stems from the kinetic energy in the field
(Armendariz-Picon et al. 2001, 2000, 1999). These
models have non-canonical kinetic energy terms. 

In addition, there are tachyonic fields, which have been a candidate
for both inflation and dark energy. For further discussion, we refer
the reader to Copeland et al. 2006, Padmanabhan 2002, Bagla et al. 2003, and
Abramo \& Finelli 2003; the most notable feature of the tachyon models is that
the equation of state varies between $0$ and $-1$ regardless of
how steep the tachyon potential is; for accelerated expansion, however,
one must have a shallow potential compared to $V(\phi)\propto\phi^{-2}$.

The models enumerated above all correspond to $w\geq-1$; however,
$w<-1$ has not been observationally disallowed, and this region of
parameter-space is generally produced by so-called {}``phantom''
or {}``ghost'' dark energy. These are scalar fields with negative
kinetic energy, and suffer from a number of difficulties, most notably
quantum instabilities (see Copeland et al. 2006 and references therein). Dilatonic models are attempts to stabilize phantom dark energy
(Cline et al. 2004, Gasperini et al. 2002). 

Finally, there is the
Chaplygin gas, with equation of state $p\propto-\rho^{-1}$, which
at early times behaves as a pressureless dust and at late times as
a cosmological constant (Copeland et al. 2006). It is thought that
the Chaplygin gas would result in a strong Integrated Sachs-Wolfe
effect, and hence loss of power in CMB anisotropies; though this can
be avoided, the constraints are rather tight (Kamenshchik et al. 2001).

\subsection{Inflation}

Since our model is based on the idea
that inflation and dark energy are the results of the same physical
process,  we briefly review the evidence
for inflation here to contextualize our model. Proposed by Guth in 1981, inflation, the idea
that the Universe underwent a period of exponential, super-luminal
expansion in the first $10^{-35}$ seconds after the Big Bang, explains
many {}``puzzles.'' For instance, the isotropy of the CMB down to
one part in $100,000$, even comparing regions that are not causally connected
in the standard Big Bang model, is a natural feature of inflation
(see Komatsu et al. 2010, Guth 2007).

Further, the Universe is flat
down to a few percent (see Tegmark et al. 2004); Dicke and Peebles
(1979) point out that to realize this flatness today in the standard Big Bang model, $\Omega_{tot}$
would have had to be unity to 59 decimal places at the Planck time,
an extraordinary fine-tuning. But in inflation, the entire observable
Universe today came from such a small initial region in comparison
to the radius of curvature of the early Universe that flatness arises
generically and requires no fine-tuning. 

Another piece of evidence
for inflation is the absence of magnetic monopoles. All Grand
Unified Theories predict them. Indeed, Preskill (1979) estimates they would be
dominant in a non-inflationary cosmology. However, none are observed.  Inflation
provides a natural mechanism to explain why, as the exponential expansion
moves the nearest monopoles beyond our observable Universe. Finally,
inflation predicts a density spectrum of almost scale-invariant, adiabatic
Gaussian fluctuations, a spectrum that fits the observed temperature
fluctuations in the CMB as a function of angular scale quite well
(Guth \& Kaiser 2005, Guth 2007, Komatsu et al. 2010). 

Thus it appears
we experienced an epoch of accelerated expansion in the early Universe
(inflation), and we observe that we are in another epoch of accelerated
expansion today due to dark energy. With all of this in mind, it seems
the simplest model of dark energy would be one in which it and the
inflation we encounter in the early Universe are essentially identical.

\subsection{Unification Models}

Unification of dark energy and inflation has been proposed before.
For instance, Peebles and Vilenkin (1998) used a scalar field potential
$V\propto\phi^{4}+m^{4},\ \phi<0,\ V\propto\frac{m^{8}}{\phi^{4}+m^{4}},\ \phi>0,$
which is supposed to produce inflation like a self-interacting field
for $\phi<0$ and then dark energy as it slow-rolls for $\phi>0$.
Cardenas (2006) proposed a tachyonic field with an exponential potential
$V\propto\exp[-\phi/\phi_{0}],$ where there is inflation followed
by reheating and finally an epoch of dark energy inflation as $\phi$
continues to roll down the hill of its potential. Finally,
Liddle \& Urena-Lopez (2006) explore two scenarios: unification of
dark matter, dark energy, and inflation into the same scalar field,
and unification of dark energy with dark matter in one scalar field
while inflation is provided by another. 

Double inflation scenarios in which the early Universe undergoes two
separate epochs of inflation have also been proposed (these scenarios
do not attempt to explain dark energy as coming from an inflaton-like
field). For instance, Poletti (1989) considers the quantum cosmology
of a mini-superspace model, and focuses on the idea that the Wheeler-DeWitt
equation may be invariant under rescalings of the metric. We note
this work simply to give credit to Poletti for the idea of a double-inflationary
scenario (he also derives the relations showing how the two fields'
evolutions relates to their masses); his focus on the Wheeler-DeWitt
equation is very different from our own on dark
energy. 

Turner et al. (1987) suggest double inflation in the early
Universe as an exlanation for a bimodal density fluctuation spectrum
(fluctuations on small and large scales) consistent with the observations
available in the late 1980's on galaxy structure. Silk \& Turner (1987)
make a similar suggestion, arguing that the decoupling of large and
small scale structure such a scenario would produce could explain
the (circa 1986) excess of observed power on large scales. Muller
and Schmidt (1988) propose that double inflation might arise from
fourth-order gravity coupled to a scalar field, and suggest that the
second epoch of inflation would yield dark matter. 

Finally, Polarski
\& Starobinsky (1992) calculate the spectrum of perturbations produced
by double inflation, and are again motivated by the apparent mismatch
between theory and observation in prediction of large-scale structure. More recently, Yamaguchi (2001) uses double inflation
to explain large-scale structure (first epoch) and small-scale structure
(second-epoch), suggesting it might naturally arise from supergravity;
Yamaguchi also does not suggest double inflation as an explanation
for dark energy.

\section{Chaotic Inflation}
\label{s:chaoticinf}

Let us begin with Linde\textquoteright{}s (1983) chaotic inflation,
arguably the simplest model of inflation ever proposed. Linde's potential
was of the form

\begin{equation}
V(\phi)=\frac{1}{2}m^{2}\phi^{2}\end{equation}

in Planck units where $c=1=\hbar=8\pi G$, $m_{pl}=(\hbar c/8\pi G)^{1/2}=1$
is the reduced Planck mass, and $G$ is Newton's constant.

This is a simple massive scalar field. The full equation of motion
for the field is

\begin{equation}
\ddot{\phi}+3H\dot{\phi}=-\frac{dV}{d\phi}=-m^{2}\phi,\end{equation}

where $H$ is the Hubble constant and $3H\dot{\phi}$ is a frictional
term due to the expansion of the Universe (see Copeland et al. 2006, Linde 2002, Lyth \& Liddle 2000). In general, in inflationary
models where the Universe is effectively flat,

\begin{equation}
H^{2}=\frac{1}{3}\rho\end{equation}

(with $8\pi G=1$).
Taking $\rho$ to be due primarily to the inflationary potential $V(\phi)$,
we have 

\begin{equation}
H=\sqrt{\frac{1}{3}}\rho^{1/2}\approx\frac{m\phi}{\sqrt{6}},\end{equation}
The approximation follows by noting that in slow-roll inflation,
the condition $\dot{\phi}^{2}<<V(\phi)$ is satisfied (see Copeland et al. 2006, Linde 2002).  Thus $\rho=\frac{1}{2}\dot{\phi}^{2}+V(\phi)\approx V(\phi)$.

We now make the further approximation that $\ddot{\phi}<<\dot{\phi,}$
which is satisfied because the frictional term $3H\dot{\phi}$ means
that the field quickly reaches a nearly constant {}``terminal''
velocity. We thus have from equation (2) that, in the early Universe
during the inflationary epoch,

\begin{equation}
\dot{\phi}\approx-\frac{m^{2}\phi}{3H}\end{equation}

Since $H= d\ln{a}/ dt$, we have

\begin{equation}
d\ln a=H\frac{dt}{d\phi}d\phi=-\frac{1}{2}\phi d\phi,\end{equation}

where the last equality follows by using equation (4) for $H$ and
equation (5) for $dt/d\phi=\dot{\phi}^{-1}.$

If $N$ is the number of e-folds of inflation then 

\begin{equation}
N=\ln\frac{a_{final}}{ a_{initial}}=-\frac{1}{2}\int\phi d\phi=\frac{1}{4}[\phi_{initial}^{2}-\phi_{final}^{2}].\end{equation}

Inflation continues in slow-roll until $\phi_{final}^{2}=1$,
when the kinetic energy $\frac{}{}$$\frac{1}{2}\dot{\phi}^{2}$
becomes comparable with the potential energy $\frac{1}{2}m^{2}\phi^{2}$.
At this point, the exponential expansion ends and the kinetic energy
in the field is dumped into the thermal energy of particles, ushering
 in the hot big bang epoch $-$ a radiation dominated, thermal-energy
filled Universe. The $\phi$ field is damped and settles at $\phi=0$
during this process, so that $V(\phi)=0$ and the vacuum energy density
in the massive scalar field becomes zero. (Or the tiny value of $V_{0}=10^{-120}$
if one adds a tiny constant to the formula such that $V(\phi)=V_{0}+\frac{1}{2}m^{2}\phi^{2}$
to account for dark energy. But we will not be adding this $V_{0}$
term.) 

Fluctuations re-entering the causal horizon today left the
causal horizon approximately $N=60$ e-folds prior to the end of inflation,
when according to equation (7) the value of $\phi^{2}=4N+1=241.$
In this case, the value of the power law primordial tilt evaluated
at $k_{0}=.002\, Mpc^{-1}$ should be 

\begin{equation}
n_{s}\approx1+2\left(\frac{V''}{V}\right)-3\left(\frac{V'}{V}\right)^{2}\approx1-\frac{8}{\phi^{2}}\approx.967\quad(predicted)
\end{equation}

(cf. Easther and McAllister 2006) and the value of $r$, the ratio
of tensor to scalar modes, should be

\begin{equation}
r\approx8\left(\frac{V'}{V}\right)^{2}\approx\frac{32}{\phi^{2}}\approx\frac{8}{N}\approx.13\quad(predicted)
\end{equation}

(cf. Kim and Liddle 2006). Fitting the amplitude of the observed fluctuations
requires

\begin{equation}
m=7.8\times10^{-6}\end{equation}

(cf. Kim and Liddle 2006). Remarkably, the predicted values of $n_{s}$
and $r$ are consistent with the observed values from WMAP+BAO+$H_{0}$
(Komatsu et al. 2010):

\begin{equation}
n_{s}=.968\pm.012\quad(observed)
\end{equation}

and the $ $$95\%$ confidence level constraint

\begin{equation}
r<.24\quad(observed).
\end{equation}

The agreement between the predicted and observed values of $n_{s}$
is especially impressive considering that potentials of the form $V(\phi)=(\lambda/4)\phi^{4}$
have been ruled out by predicting unacceptable values of $n_{s}=.95$
and $r=.26$ (cf. Komatsu et al. 2010). Given this, the search for
the tensor modes $(r>0)$ is on $-$ for instance, the $Planck$ satellite
hopes to improve the measurement of $r$. Polarization studies in
the future should if successful offer a smoking-gun proof that the
tensor modes are there. Such modes are not predicted by the Ekpyrotic/Cyclic
scenario and, if found, they would offer a convincing proof of inflation
(cf. Linde 2002). 

It is remarkable that a model as simple as Linde\textquoteright{}s
massive-scalar-field chaotic inflation is currently consistent with
the observational data. Linde's model predicts in particular a value
of $n_{s}$ noticeably less than $1$ $-$ a hallmark of slow-roll
inflation that is indeed observed. 

While more complicated potentials can produce lower values of $r$
(cf. Kallosh and Linde 2010), this comes at the expense of adding
more free parameters. The Linde theory, in contrast, is simple enough
that it offers the possibility of being confirmed in a dramatic way
if the observed value of $r$ is $.13$. If that occurs, we will no
doubt conclude that the inflation seen in the early Universe is due
to a massive scalar field. In that case, we argue here that we should
expect a similar origin for dark energy as well. 

\section{Double Inflation}
\subsection{$\delta w_0$ in terms of the dark energy scalar field today}
\label{s:deltawtoday}

We propose the following simple double-inflation potential for inflation
and dark energy:

\begin{equation}
V=\frac{1}{2}m_{1}^{2}\phi_{1}^{2}+\frac{1}{2}m_{2}^{2}\phi_{2}^{2}\end{equation}

with $m_{1}\sim10^{-5}$ and $m_{2}\leq10^{-60}.$ Double inflation
with a potential of this form was introduced (typically with $m_{1}\sim(20\pm5)m_{2}$)
to explain inflation alone (Polarski \& Starobinsky 1992, see also
Silk \& Turner 1987). We will be using it with widely different mass
scales to explain inflation and dark energy. The equation of motion
for the inflaton field is

\begin{equation}
\ddot{\phi}_{1}+3H\dot{\phi_{1}}+m^{2}\phi_{1}=0,\end{equation}

and for the dark energy field is

\begin{equation}
\ddot{\phi}_{2}+3H\dot{\phi_{2}}+m^{2}\phi_{2}=0.
\label{eom}
\end{equation}

As noted earlier, we are working in Planck units where $c=1=\hbar=8\pi G$, $m_{pl}=(\hbar c/8\pi G)^{1/2}=1$
is the reduced Planck mass, and $G$ is Newton's constant. 

In the inflationary epoch when the Universe is dominated by the vacuum
energy density provided by $V$, we have

\begin{equation}
3H^{2}=V=\frac{1}{2}m_{1}^{2}\phi_{1}^{2}+\frac{1}{2}m_{2}^{2}\phi_{2}^{2}.
\end{equation}

The slow-roll approximation is valid for both fields and the evolution
of the two fields is given by 

\begin{equation}
\ln\frac{\phi_{2}(t)}{\phi_{2,initial}}=\left(\frac{m_{2}^{2}}{m_{1}^{2}}\right)\ln\frac{\phi_{1}(t)}{\phi_{1,initial}}\end{equation}

(cf. Easther \& Mcallister 2006). Since $m_{2}<<m_{1,}$ $\phi_{2}(t)\approx constant\approx\phi_{2,initial}$
even though $\phi_{1}(t)/\phi_{1,initial}$ evolves considerably during
inflation. There have been $60$ e-folds of inflation since the perturbations
now re-entering the horizon left the causal horizon, but there could
have been more e-folds of inflation before that, so we expect $N>60$
and thus $\frac{}{}$$\frac{1}{4}[\phi_{1,initial}^{2}-\phi_{1,final}^{2}]=N>60.$
Since $\phi_{1,final}^{2}=1$ marks the end of inflation, $\phi_{1,initial}^{2}>(4\times60+1)=241.$
As inflation ends, the kinetic energy in the $\phi_{1}$ field is
converted into thermal particles, the motion in $\phi_{1}$ damps,
and $\phi_{1}$ comes to rest at a value of $\phi_{1}=0.$ The dark
energy seen today thus derives from the $\phi_{2}$ field, and $V=\frac{1}{2}m_{2}^{2}\phi_{2}^{2}$ today because $\phi_{1,0}=0$.

Since we observe a dark energy acceleration today consistent with
$w\approx-1$ we expect slow-roll inflation to apply $\left[\ddot{|\phi|}<<3H|\dot{\phi|},\ \dot{\phi}^{2}<<V(\phi)\right]$.
Since dark energy is dominant today, we use our results from Section \ref{s:chaoticinf}.  Since today $\rho_{DE,0}=\Omega_{DE}\rho_0$,
$\rho_0=(1/\Omega_{DE})\rho_{DE,0}.$ Making these changes to equation
(4), we have 

\begin{equation}
H_0\approx\frac{1}{\sqrt{6\Omega_{DE}}}m_{2}\phi_{2,0},\; \dot{\phi}_{2,0}\approx-\frac{m_{2}^{2}\phi_{2,0}}{3H_0}\approx-\sqrt{\frac{2\Omega_{DE}}{3}}m_{2}.
\label{Handphi}
\end{equation}
Now, $w$, defined to be the ratio of pressure to energy density in the dark energy, is given by

\begin{equation}
w\equiv\frac{p}{\rho}=\frac{\frac{1}{2}\dot{\phi}_{2}^{2}-V}{\frac{1}{2}\dot{\phi}_{2}^{2}+V},
\label{weq}
\end{equation}

where the equality is specific to scalar quintessence
fields (see Copeland et al. 2006), and we recall that because the inflationary field has damped long ago, $V=\frac{1}{2}m_2^2\phi_2^2$ in the present epoch.  We therefore have from equations (\ref{Handphi}) and (\ref{weq}) that

\begin{equation}
w_0\approx\frac{\frac{2}{3}\Omega_{DE}-\phi_{2,0}^{2}}{\frac{2}{3}\Omega_{DE}+\phi_{2,0}^{2}}.\end{equation}

The difference between $w_0$ and $-1$, which we define to be $\delta w_0$, is thus

\begin{equation}
\delta w_0\equiv w_0+1\approx\frac{4}{(3/\Omega_{DE})\phi_{2,0}^{2}+2}.
\label{boe}
\end{equation}

Current limits from WMAP suggest $\delta w_{0}=.07\pm.13,$ where
$\delta w_{0}$ is simply $\delta w$ today. Importantly, since in the future
we expect $\phi_{2}$ to roll down to zero, leaving a vacuum energy
density of zero (with no $V_{0}$ term), the Boltzmann brain problem
disappears.  See Figure \ref{fig1} for a comparison of this formula to our numerical results from a full and self-consistent solution of the equation of motion for $\phi_2$ and the Friedmann equation as described in Section \ref{method}.  Equation (\ref{boe}) is only valid evaluated today, because today we are dark-energy dominated, which is the assumption under which this formula was derived.  This formula will not be valid as a function of time in the past, because dark energy has only recently become dominant.

\subsection{$\delta$w as a function of redshift}
 In what follows, we will derive a formula that gives  $\delta w$ as a function of time. Let us be very clear here that we are no longer assuming dark-energy domination as we did in \ref{s:deltawtoday}. 
 
 The formula we present below will give $\delta w(z)$ as a function of $H(z)$ and so is a simple template observational programs can use to look for slow-roll dark energy.

We have from equation (\ref{weq}) and the definition $\delta w=w+1$ that

\begin{equation}
\delta w=\frac{\dot{\phi}_{2}^{2}}{(1/2)\dot{\phi}_{2}^{2}+V}.
\end{equation}
From this, we can see that small $\delta w$ implies that $\dot{\phi}^2<<V$, and so we approximate that
\begin{equation}
\delta w\approx\frac{\dot{\phi}_{2}^{2}}{V}.
\end{equation}
Since $ $$\dot{\phi}_{2}$ is small, $\phi_{2}$ does not vary much
with time and so $V$ is approximately constant in time.  Thus $\delta w\propto\dot{\phi}_{2}^{2}.$
Since in slow-roll $\ddot{\phi}_{2}$ is also small, we have from
equation (\ref{eom}) that \begin{equation}
3H\dot{\phi}_{2}\approx - \frac{\partial V}{\partial\phi_{2}}=-m_{2}^{2}\phi_{2}.\end{equation}

As we have already noted, $\phi_{2}$ is approximately constant. Hence $3H\dot{\phi}_{2}\approx constant$,
so 

\begin{equation}
\dot{\phi}_{2}\propto\frac{1}{H}.
\end{equation}

Since $\delta w\propto\dot{\phi}_{2}^{2}$, equation (25) implies that
\begin{equation}
\delta w\propto\frac{1}{H^{2}}.\end{equation}

Normalizing appropriately, we obtain

\begin{equation}
\delta w(z)\approx\delta w_{0}\left(\frac{H_{0}}{H(z)}\right)^{2}.
\label{slowroll}
\end{equation}

In summary, then, for small $\delta w$, slow-roll applies, $\dot{\phi}_{2}$
is small, and $\delta w\propto\dot{\phi}_{2}^{2}.$ When the field
reaches terminal velocity, the acceleration $\ddot{\phi}_{2}$ is
nearly zero, so the Hubble friction term $3H\dot{\phi}_{2}$ is balanced
by the slope of the potential the field is rolling down, $-\partial V/\partial\phi_{2}$.
This is analogous to a ball rolling down a hill: it will reach terminal
velocity when energy dissipation from friction cancels out the energy
it gains by moving to lower values of its potential. Finally, if $\phi_{2}$
is roughly constant in time, $\partial V/\partial\phi_{2}$ will be
as well, meaning $\dot{\phi}_2\propto H^{-1}$, which leads to equation (26). This derivation
could easily be duplicated for other dark energy potentials as long
as $\delta w$ is small.

We term this formula (equation (\ref{slowroll})) our
"slow-roll formula'' in subsequent discussion.  See Figure \ref{Hfitslowroll} for a comparison of this formula to our numerical results from a full and self-consistent solution of the equation of motion for $\phi_2$ and the Friedmann equation as described in Section \ref{method}.

\section{N-flation}

A potential of the form in equation (13) governed by equations of
motion (14) and (15) can be produced easily by models of N-flation.
A possible criticism of the original Linde chaotic inflation is that
it requires $\phi>1$ in Planck units. Linde argued that this
was acceptable as long as $V<1.$ But it was felt that it would be difficult
to produce values of $\phi>1$ in string theory models. Thus, N-flation
(Dimopoulos et al. 2008) has been proposed (cf. also Easther and McAllister
2006), motivated by the fact that supersymmetric string theories allow of order $10^{5}$ axion
fields. 

For such axion fields $V(\psi_{i})=\mu^{4}[1-\cos(\psi_{i}/f_{i})]$,
and for $\psi_{i}<<1$ (i.e. $\psi_{i}$ significantly below the Planck
mass $-$ which we would like) we note that the potential is of the Linde
quadratic form with effective mass $m=\mu^{2}/f_{i}$. As we have
already pointed out, string theory allows the number of such fields
to be large (Dimopoulos et al. 2005). Hence in this work we will adopt $N=10^{4}$.

In this model there are $N$ fields $\psi_{i}$ (where $i=1,\,.\,.\,.\,,\, N$)
with approximately equal masses $m_{i}\approx m$. Then the potential
is  $V=\sum\frac{1}{2}m_{i}^{2}\psi_{i}^{2}\approx V(\phi)=\frac{1}{2}m^{2}\phi^{2}$ where
 $\phi^{2}\equiv\sum\psi_{i}^{2}$. This is so-called "assisted
inflation." 

Since each $m_{i}\approx m$, all the $\psi_{i}$'s evolve together
via
 \begin{equation}
\ln\left[\frac{\psi_{i}(t)}{\psi_{i,initial}}\right]=\left(\frac{m_{i}^{2}}{m_{j}^{2}}\right)\ln\left[\frac{\psi_{j}(t)}{\psi_{j,initial}}\right]\approx\ln\left[\frac{\psi_{j}(t)}{\psi_{j,initial}}\right]
\end{equation}

 if $m_{i}\approx m_{j}$ for all $i$ and $j.$ This will be true if
the mass spectrum of the fields is strongly peaked and densely
packed:

\begin{equation}
m_{i}^{2}=m^{2}\exp\left[(i-1)/\sigma\right],\end{equation}

where $\sigma>280$ for $N>600$ (Kim and Liddle 2006). If the fields
are strictly non-interacting the masses could in fact be exactly equal. 

We want double N-flation with a potential 

\begin{equation}
V=\sum\frac{1}{2}m_{1}^{2}\psi_{1,i}^{2}+\sum\frac{1}{2}m_{2}^{2}\psi_{2,i}^{2}=\frac{1}{2}m_{1}^{2}\phi_{1}^{2}+\frac{1}{2}m_{2}^{2}\phi_{2}^{2}
\label{nflation}
\end{equation}
where there are $N=10^{4}$ $\psi_{2}$ fields each of mass $m_{2}\approx10^{-60}$
and $N=10^{4}$ $\psi_{1}$ fields each of mass $m_{1}\approx10^{-5}$
(in keeping with our hypothesis that inflation and dark energy should
arise from the same process) and by definition $\phi_{2}^{2}=\sum\psi_{2,i}^{2}$
and analogously for $\phi_{1}^{2}.$ Initially we need $\phi_{1,initial}^{2}>241$
to produce enough inflation ($>60$ e-folds) to explain our Universe.
With $10^{4}$ fields, that just means that each $\psi_{1,i}^{2}>.0241$
and so all the $\psi_{1,i}$'s can be sub-Planckian ($<1$). This
is good. 

Are such low values of $m_{2}\approx10^{-60}$ plausible from the
point of view of string theory? Interestingly, Kaloper and Sorbo (2006)
have independently proposed just such an N field quiNtessence model
for dark energy using ultralight pNGB {[}pseudo-Nambu Goldstone bosons{]}
(axions) from string theory. They argue for potentials of the form
$V(\psi_{i})=\mu^{4}[1-\cos(\psi_{i}/f_{i})]$. Svrcek (2006) has
also argued for multiple ultra-light axion fields with potentials
of this form to explain dark energy. 

We note in each case that for sub-Planckian $\psi_{i,initial}$'s,
the potential is of the desired Linde quadratic form with effective
mass $m=\mu^{2}/f_{i}$. Svrcek notes that pseudoscalar axion fields
have a shift symmetry and if this symmetry were exact it would set
the potential to zero and the axions would be massless. In string
theory the shift symmetry is broken only by nonperturbative effects.
In string theory axions thus receive potential only from nonperturbative
instanton effects which are exponentially suppressed by the instanton
action. Hence, if the instantons have large actions they can give
rise to a potential many orders of magnitude below the Planck scale. 

Svrcek argues that $\mu^{4}=M^{4}\exp[-S_{inst}],$ where $M\sim1$
and $S_{inst}\sim280,$ can create a vacuum energy density today comparable
with what we observe for dark energy. Svrcek adds a $V_{0}$ term
as well, which we eliminate as unnecessary. We argue that if
the axion fields are able to explain the amount of dark energy we
observe today the $V_{0}$ term can be eliminated. Both Kaloper and
Sorbo (2006) and Svrcek (2006) are explicitly creating quintessence
models for dark energy. Both also note that single field models with
sub-Planckian field values are unacceptable for quintessence and favor
models with $N=10^{4-5}$ fields. 

We are proposing to combine these quintessence models that use $N$
fields with the N-flation models to explain dark energy $and$ inflation.
Independently Svrcek (2006) also speculates \textquotedblleft{}Hence,
it could be that some of the string theory axions have driven inflation
while others are currently responsible for {[}a{]} cosmological constant.\textquotedblright{}
We take the point of view here that there are $N=10^{4}$ equal-mass
ultra-light fields $\psi_{2,i}$ that create a slow-roll dark energy
and $N=10^{4}$ equal-mass heavy fields $\psi_{1,i}$ that create
slow-roll inflation in the early Universe. 

If there are in addition singleton fields with intermediate masses,
with sub-Planckian $\psi_{i}$'s also, these would not have inflated
but rather would have rolled down, as Svrcek notes. Some of these axion fields could
have rolled down and created dark matter. They do not cause inflation
because $\psi_{i}<1$, so when the other thermal particles redshift
so that the $\psi_{i}$ vacuum field energy becomes dominant, $3H^{2}=\frac{1}{2}m_{i}^{2}\psi_{i}^{2}$.
Thus $H$ is not large enough to cause the low velocity ($\dot{\psi_{i})}$
required for slow-roll inflation. 

However, if there are many fields of essentially the same mass, $3H^{2}=\sum\frac{1}{2}m_{i}^{2}\psi_{i}^{2}\approx\frac{1}{2}m^{2}\phi^{2},$
where $\phi^{2}>1$. In this case, $H$ is much higher (by a factor
of $\sqrt{N}$) causing each $\dot{\psi}_{i}$ to be lower and slow-roll
inflation to occur. Thus, inflation only occurs when many fields congregate
at the same mass scale. 

\section{Why N-flation is superior}

We expect all of the $\psi_{2,i}$'s and $\psi_{1,i}$'s to be sub-Planckian.
If $N=10^{4}$, this means that $\phi_{i,initial}^{2}<10^{4}$, which
means (using equation 7) that there can be at most 2500 e-folds of
inflation in our Universe. Hence our Universe today is less than $\exp[2440]$
times larger than the visible horizon. In Linde\textquoteright{}s
original formulation of chaotic inflation, random quantum fluctuations
allowed Universes to give birth to Universes with various values of
$\phi_{1,initial}$. The ones with larger values of $\phi_{1,initial}$
grew faster until most of the volume of the multiverse was in the
fastest expanding states, with $V=1=\frac{1}{2}m_{1}^{2}\phi_{1}^{2}.$
That would mean $\phi_{1,initial}^{2}\approx10^{10}$ and the Universe
today would be $\exp[10^{10}]$ times larger than the part we can
see. 

For our model of dark energy, a simple Linde-type chaotic double-inflation
picture would eventually lead to most of the volume of the multiverse
being in the fastest expanding states, given by the ellipse 

\begin{eqnarray}
V=1=\frac{1}{2}m_{1}^{2}\phi_{1}^{2}+\frac{1}{2}m_{2}^{2}\phi_{2}^{2} \nonumber \\
=\frac{\phi_{1}^{2}}{2\times\left(10^{5}\right)^{2}}+\frac{\phi_{2}^{2}}{2\times\left(10^{60}\right)^{2}}.
\end{eqnarray}

This ellipse is very elongated in the $\phi_{2}$ direction, and at
random points on it the $\phi_{2}$ field contributes just as much
to the potential and to the inflation as the $\phi_{1}$ field. Thus starting
values $\phi_{2,initial}\approx(m_{1}/m_{2})\phi_{1,initial}$ would
be expected, and since $\phi_{2}$ is slower to roll down than $\phi_{1},$
we would not get the sub-dominant dark energy we require. 

If we use N-flation to realize the potential in equation (13) via
equation (\ref{nflation}) this problem does not occur. Since all of the fields
are sub-Planckian, the fastest inflating regions are characterized
by starting values of $\phi_{2}^{2}\approx\phi_{1}^{2}$ that are
bounded above by $N$, and since $m_{2}<<m_{1}$, the contribution
of the $\phi_{2}$ field to the potential is sub-dominant ($<\frac{1}{2}m_{2}^{2}N$
for the $\phi_{2}$ field versus $<\frac{1}{2}m_{1}^{2}N$ for the
$\phi_{1}$ field). 

\section{Bound on probabilities for deviation of \lowercase{w} from $-1$}

In this picture we might expect the initial values for $\phi_{2}^{2}$$ $
and $\phi_{1}^{2}$ to be comparable. What is the smallest $\phi_{1,initial}^{2}$
could be? It must be at least 241 to explain the at least 60 e-folds of inflation
we see within the visible Universe. By the above argument, we might
expect $\phi_{2,initial}^{2}$ to be similar. Because of equation
(17), $\phi_{2}$ does not evolve much during the period of inflation.
Its main chance to roll down is in the current epoch when $H$ is
low. But there have not been many e-folds of inflation during the
current epoch, so we might expect $\phi_{2,0}^{2}$ to be only a little
less than its minimal initial value of about $241$. That would give a value
today of $\delta w_{0}\approx4/\left(\left(3\phi_{2,0}^{2}/\Omega_{DE}\right)+2\right)=.4\%$ via equation (21). 

Since we expect the $\psi_{2,i}$'s and the $\psi_{1,i}$'s of equation (30) to be
uncorrelated in our n-flationary picture, we expect by the central limit theorem that $\phi_{2,initial}$
and $\phi_{1,initial}$ are Gaussian. If their magnitudes are comparable,
in keeping with our hypothesis that the physical processes for inflation
and dark energy are identical, we might a priori expect on average
to find $<\phi_{2,initial}^{2}>\:=\:<\phi_{1,initial}^{2}>.$ 

Consider the probability function $P(\phi_{2,initial},\phi_{1,initial})d\phi_{2,initial}d\phi_{1,initial}.$
If both variables follow a Gaussian distribution, $P$ has
circular probability contours in the $(\phi_{2,initial},\,\phi_{1,initial})$
plane, and the probability of finding $|\phi_{2,initial}|<X|\phi_{1,initial}|$
for $X<1$ is simply $P=(2/\pi)\arctan X$. If $\phi_{1,initial}^{2}>241,$
the probability of observing a value of $\delta w_{0}>\delta w_0'$ is $ $

\begin{equation}
P(\delta w_0 > \delta w_0') <\frac{2}{\pi}\arctan\left[\frac{1}{\sqrt{241}}\sqrt{\frac{\Omega_{DE}}{3}\left[\frac{4}{\delta w_0'}-2\right]}\right]
\label{prob}
\end{equation}

using equation (21) to relate $\phi_{2,0}\simeq \phi_{2,initial}$ to $\delta w_{0}$. See Fig. 1.  The probability
of observing $\delta w_{0}>3\%$ is $P<20\%.$ It could be considerably
smaller if there are significantly more than the minimum 60 e-folds
of inflation in the early Universe.

\begin{figure}
 \includegraphics[width=84mm]{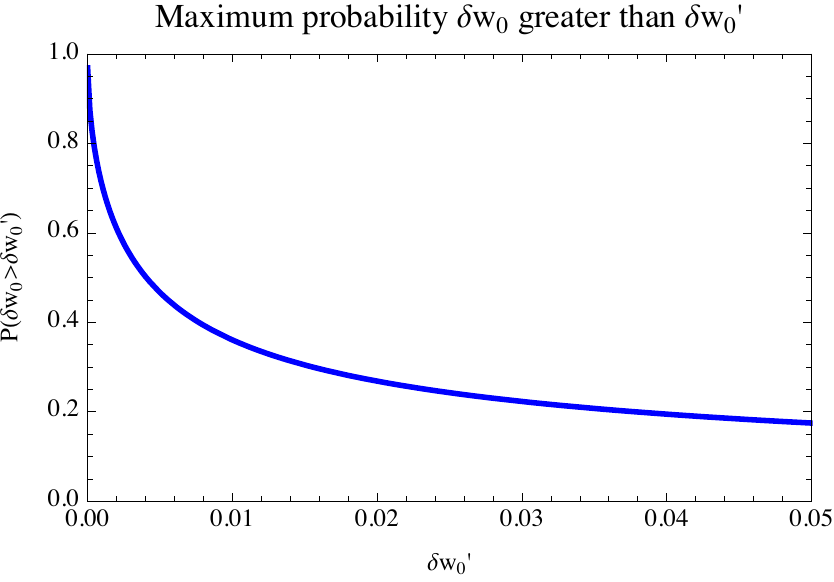}
\centering{}\caption{Plot of our upper bound on the probability that $\delta w_{0}$ is greater
than a given value (equation (\ref{prob})).  This is an upper bound because in fact $\phi_{1,initial}$ could be much greater
than $16$ (chosen to provide approximately the minimal 60 e-folds
of inflation), leading to larger values of $\phi_{2}$ and lower values of $\delta w_0$.  However, the
plot is moderately encouraging: there still could in principle be
a nearly $20\%$ chance that $\delta w_{0}$ could be as large as
$3\%$, which would likely be detectable.\vspace{0.5in}
}

\end{figure}

With this in mind, it may be reasonable to expect a small value of $\delta w_{0}$.
On the assumption that the same physical processes led to both inflation
and dark energy, if there were at least $60$ e-folds of inflation
in the early Universe, there should be at least $60$ e-folds of inflation
due to dark energy ahead in the future, as indicated in the calculation
above. However, the number of galaxies (and observers like ourselves)
produced in our Universe is proportional to 

\begin{equation}
\exp[3N]=\exp\left[\frac{3}{4}\left(\phi_{1,initial}^{2}-1\right)\right].\end{equation}

So it might not be surprising for us to observe 
\newline $<\phi_{2,initial}^{2}>$
less than $<\phi_{1,initial}^{2}>$, but the details depend on questions
of measure which are unsettled. 

Suffice it to say, one must find a measure that makes what we observe,
namely $\phi_{2,initial}^{2}\gtrsim12$ (since $\delta w_{0}=7\%\pm13\%$
from WMAP) and $\phi_{1,initial}^{2}>241$ (to produce at least $60$
e-folds of inflation), not particularly unlikely. For the time being,
we suggest taking an empirical approach and asking what future observations
can tell us about $\phi_{2,initial.}$

\section{Numerical Method}
\label{method}

To integrate the equation of motion for $\phi_{2}$ (15) numerically, we must also
solve the Friedmann equation for $H$ as a function
of time. Since we know from observational constraints
that $\delta w$ is small, we start by assuming that $w=-1$ and solve
the Friedmann equation to determine $H$ as a function of time. Then
using this $H$ we can solve the full equation of motion for $\phi_{2}$
as a function of time (though we will instead solve the re-scaled
equation (43) as described later in this section). Knowing $\dot{\phi}_{2}$
gives us $w$ as a function of time, and we can insert this back into
the Friedmann equation to give us an improved $H$ as a function of time. 

We then iterate.
This approach converges rapidly on a self-consistent solution where
we have $w,\ H,$ and $\phi_{2}$ as functions of time and both the Friedmann equations
and the equation of motion for the field $\phi_{2}$ have been solved
self-consistently and exactly. 

For starting conditions at $t=0$ we assume $d\phi_{2}/dt$=0
since $H(t)$ tends to infinity as $t$ tends to zero. We discuss
the starting condition on $\phi_{2}$ itself later. Since in this
section we will consider only the $\phi_{2}$ field with mass $m_{2},$
for greater legibility we suppress the subscripts on $\phi_{2}$ and
$m_{2}$ in what follows. We recall equation (15) (with subscripts suppressed):

\begin{equation}
\ddot{\phi}+3H\dot{\phi}+m^{2}\phi=0,
\label{eomrecall}
\end{equation}

where $H=(1/a)(da/dt)$ is the Hubble constant. Defining
a new variable for time $\tau=t/t_{H0}=tH_{0}$ and  transforming $\ddot{\phi}$
and $\dot{\phi}$, we have $\ddot{\phi}=\phi''H_{0}^{2}$
and $\dot{\phi}=\phi'H_{0}$,
where prime denotes a derivative with respect to the new time variable
$\tau$. We thus have from equation (\ref{eomrecall})
that

\begin{equation}
\phi''+3\frac{H(t)}{H_{0}}\phi'+\frac{m^{2}}{H_{0}^{2}}\phi=0.\end{equation}

$\phi$ and its derivatives are functions of $\tau$, so we desire that $H$ should be as well.  We
convert the Friedmann equation from time variable $t$ to time variable
$\tau$; so doing will yield the Friedmann equation governing $H(\tau) \equiv H(t)/H_{0}$. 

It is 

\begin{eqnarray}
H^{2}(\tau) \equiv \big( \frac{1}{a} \frac{da}{d\tau} \big)^2 = \big{(} \Omega_r a^{-4}+\Omega_m a^{-3} \\
 +\;  \Omega_{DE} \exp \left[ 3 \int_{\tau}^{\tau_0} H (\tau') \left[ (1+w(\tau') \right] d\tau' \right]  \big{)}, \nonumber 
\end{eqnarray}

where $\Omega_r\equiv \rho_r/\rho_c$, and analogously for $\Omega_m$ and $\Omega_{DE}$. $\rho_c\equiv3H_0^2/(8\pi G)$ is the critical density; subscript $r$ denotes radiation, $m$ matter, and $DE$ dark energy.   $\tau_{0}$ is the value of $\tau$ today (when $a=1$). Note that for $w\equiv-1,$ this reduces to the usual Friedmann equation
used in $\Lambda CDM$ models. We thus find the
equation for $\phi$

\begin{equation}
\phi''+3H(\tau)\phi'+m_{*}^{2}\phi=0,\end{equation}

where we have defined $m_{*}^{2}=m^{2}/H_{0}^{2}$. 

In principle,
we can now solve the Friedmann equation numerically for $H(\tau)$
and using it obtain $\phi$ numerically. However, there is a constraint
on $\phi_{2,0}$ that we must satisfy as we solve numerically. The
Universe is flat with $\Omega_{DE}+\Omega_{M}\approx1$. Since dark
matter is present, the value of $H_{0}^{2}$ is larger than it would
be if only dark energy were present by a factor of $\Omega_{DE}^{-1}$;
WMAP-7 gives $\Omega_{DE}=.73$ (Komatsu et al. 2010). Thus at present

\begin{equation}
3H_{0}^{2}=\frac{1}{2\Omega_{DE}}m^{2}\phi_{0}^{2}X,
\end{equation}

where $X\equiv(\dot{\phi_{0}}^{2}+m^{2}\phi_{0}^{2})/m^{2}\phi_{0}^{2}.$

Using the definitions of $w$ and $\delta w$, we find that 

\begin{equation}
m_{*}^{2}=\frac{3\Omega_{DE}(2-\delta w_{0})}{\phi_{2,0}^{2}}.
\label{H0Fman}
\end{equation}

Hence we are not free to choose both $m_{*}$ and $\phi_{2,initial}$
independently because $\phi_{2,initial}$ will determine $\phi_{2,0}$,
which must be consistent with $m_{*}$ such that equation (\ref{H0Fman}) is
satisfied. We therefore require a method of solving the equation of
motion where we can set $\phi_{2,initial}$ (and hence $\phi_{2,0}$)
after we already have a solution. This motivates us to observe that
the equation can be dynamically rescaled by writing $\phi'/\phi=\zeta(\tau)$.
It is evident that $\phi$ is always non-zero. Writing $\phi'=\zeta\phi$,
we find that $\phi''=\left(\zeta^{2}+\zeta'\right)\phi$.
Substituting these relations into equation (37), we obtain

\begin{equation}
\zeta^{2}+\zeta'+3H(\tau)\zeta+m_{*}^{2}=0.\end{equation}

We choose $m_{*}$ and numerically solve this equation beginning at
$\tau=0$, where we have $\zeta=0$ because of our earlier comment
on $\dot{\phi}_{2,initial}.$ This is done iteratively in conjunction
with the solution of equation (36) as in the overview at the beginning
of this section, and in greater detail below; the final iteration
gives us $\delta w_{0}.$ We can then determine $\phi_{2,0}$ by inverting
equation (\ref{H0Fman}). 

We have already given an overview of our iterative procedure; here
we give details. As noted before, we begin by finding $H$ as a function
of $\tau$ for $w=-1$, which we do using the Friedmann
equation (36). For $w=-1$, this does not require knowing $\tau_{0}$.
  However, solving for $\tau_{0}$ $(\equiv\tau$ such that $a=1$) in this first
iteration does provide $\tau_{0}$ for the next round of iteration,
where we will use $w=w(\tau)$ in equation (36). We use $H$ as provided
by (36) in the dynamically rescaled equation (40) for $\zeta\equiv \phi'/\phi$,
which yields $w=(\zeta^2-m_*^2)/(\zeta^2+m_*^2)$ via equation (19). We insert this $w(\tau)$ back into (36) and find a new
$H$, which goes back into (40), again yielding $w$. We continue
this process until the difference between the $n^{th}$ and $n+1^{th}$
step integrated over all time is negligible compared to $\delta w_{0}$.
Then we have an exact and self-consistent solution to the equation
of motion for the $\phi_{2}$ field and the Friedmann equation.

\section{Numerical Results}

\begin{figure}
\includegraphics[width=84mm]{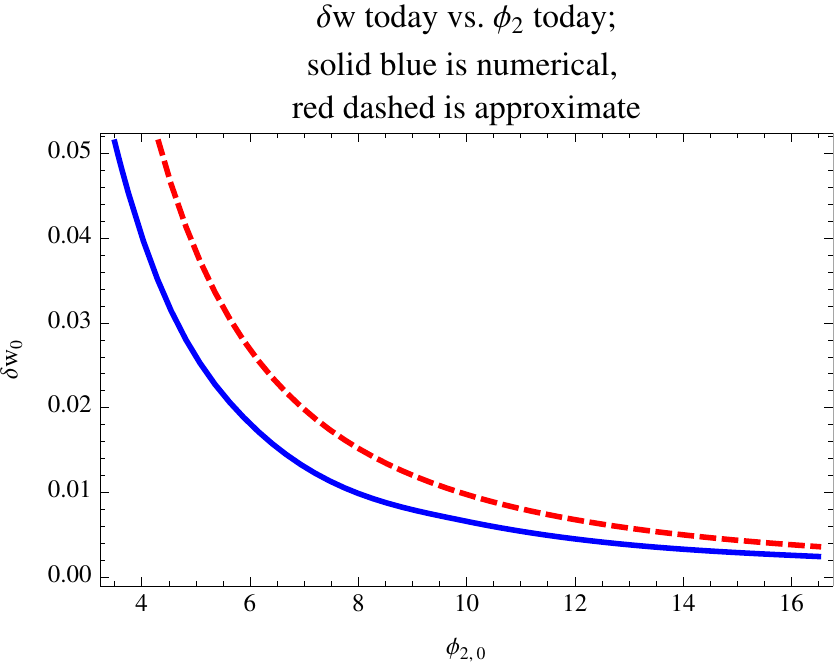}
\centering{}\caption{Plot comparing the {}``back-of-the-envelope'' approximate formula
$\delta w_{0}=4/(\frac{3}{\Omega_{DE}}\phi_{2,0}^{2}+2)$ with the
numerical results obtained as outlined in Section \ref{method}. The plot shows
that, while the back-of-the-envelope calculation captures the order-of-magnitude
of $\delta w_0$ and the rough shape of the curve, it should not be
trusted in lieu of the numerical results. Since it was derived on
the assumption that $\delta w_{0}<<1$, it makes sense that the approximate
and numerical results are closer in this region of the plot (i.e. for larger $\phi_{2,0}$). 
\vspace{0.5in}}
\label{fig1}
\end{figure}

In this section, we discuss our numerical results.  Figure \ref{fig1} compares our approximate formula for $\delta w_0$ as a function of the value of the dark energy scalar field today, (equation (\ref{boe})), to our numerical results from exact and self-consistent solution of the Friedmann and scalar field equations, described in Section 8.  The curves have similar shapes.  However, the approximate formula is not a good substitute for the full numerical results in the range $4<\phi_{2,0}<12$.  For $\phi_{2,0}>12$, the match between analytical and numerical is somewhat better. Since our approximate formula was derived in the limit that $\delta w_0 <<1$, it is not surprising that it agrees better with the full numerical results the smaller $\delta _{w,0}$ becomes.

\begin{figure}
 \includegraphics[width=84mm]{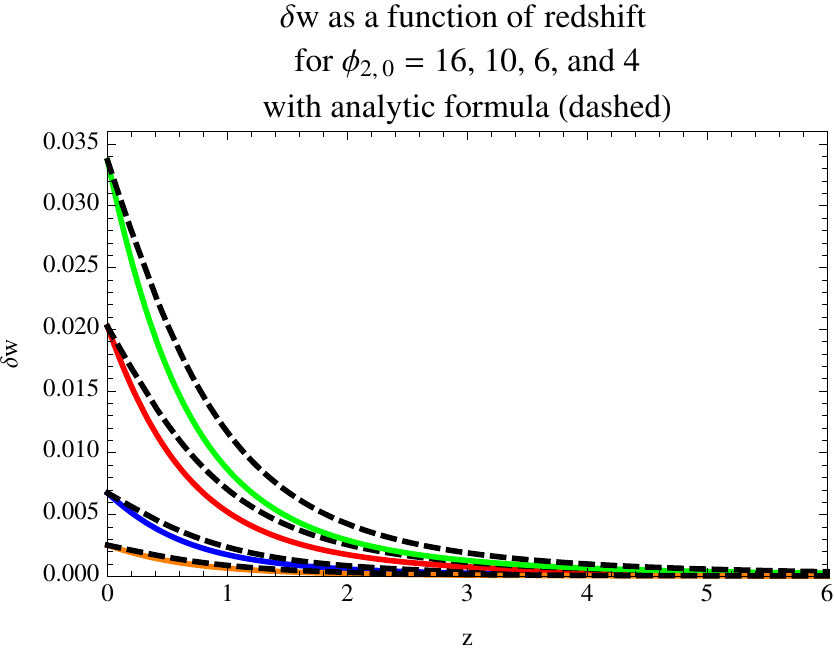}
\centering{}\caption{Here, we show our numerical results for $\delta w$ as a function
of redshift and also the "slow-roll formula" (dashed) $\delta w=\delta w_{0}\left(\frac{H_{0}}{H(z)}\right)^{2}$
(see Section 4 for derivation). The values of $\phi_{2,0}$ listed
on the plot go 16, 10, 6, and 4 from bottom to top for the solid curve/dashed curve
pairs. For smaller $\delta w_{0}$ the assumptions underlying our
"slow-roll formula" are better satisfied, and one can see it
agrees more closely with the numerical results in this regime. \vspace{0.5in}
}

  \label{Hfitslowroll}
\end{figure}

Figure \ref{Hfitslowroll} compares our "slow-roll formula" (equation (\ref{slowroll})) with our full numerical results from exact and self-consistent solution of the Friedmann and scalar field equations.  Note that the function $H$ used in these plots comes in each case from the self-consistent solution of the Friedmann equation as described in section 8. The "slow-roll" formula agrees well with our numerical results and best for $\delta w_0\lesssim 1\%$.  Analogously to our discussion of Figure \ref{fig1}, this is not surprising because the "slow roll formula" was derived in the same limit that $\delta w_0<<1$.  Hence the smaller $\delta w_0$ is, the better we would expect the agreement to be.

\begin{figure}
 \includegraphics[width=84mm]{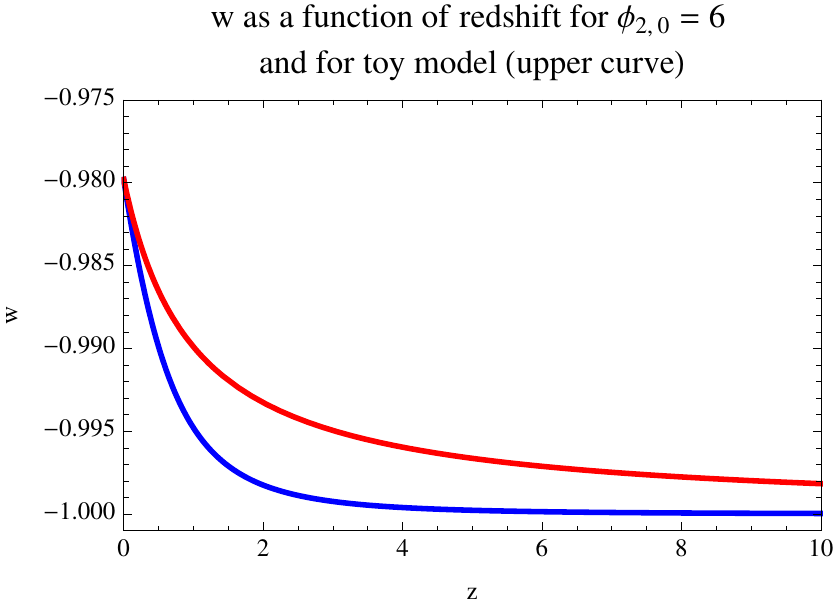}
\centering{}\caption{Here, we compare our model's predictions for $w$ as a function of
redshift to those of the popular toy model $w=w_{0}+w_{a}(1-a),$
where $w_{0}$ and $w_{a}$ are constants and $a$ is the scale factor
appearing in the Friedmann equation. The toy model, fit at both ends, is the shallower,
upper curve. The comparison shows that it will be harder to observe deviations
from $w=-1$ in the past if our model is correct than if the toy model
is correct. $ $ \vspace{0.5in}
}

 \label{toymodel}
\end{figure}

\newpage
Figure \ref{toymodel} compares a representative result from our model (with $\delta w_0=2\%$) to the Chevallier-Polarski-Linder parametrization, $w=w_0+w_a(1-a)$.  We chose the parameters $w_{0}$ and $w_{a}$ so that
the toy model agrees with ours at $a=1$ and at $a=0$ (today and
at the Big Bang). This leads to $w_{0}=-1+\delta w_{0}$ and $w_{a}=-\delta w_{0}.$  Since the CPL parametrization is the shallower curve in the plot, it is clear that if our model is correct, it will be harder to observe $w$ different from $-1$ in the past than if the CPL parametrization is right.  A comparison of Figures \ref{Hfitslowroll} and \ref{toymodel} shows that our slow-roll formula provides a much better fit for our numerical results than the CPL parametrization, as it has a physical motivation from the slow-roll approximation.

 Figure \ref{Hfitslowroll} tells a story consistent with Figure \ref{fig1}: the larger $\phi_{2,0}$ is, the smaller is $\delta w_0$.  The main take-away is that $\delta w$ is a quite steep function of redshift; as the figure shows, $\delta w$ is nearly indistinguishable from $-1$ for $z\gtrsim 3$ for all of the curves (different curves correspond to different values of $\phi_{2,0}$.)  This means that if our model is correct, it will be very challenging to observe deviations of $w$ from $-1$ except in the recent past.  If our model is correct, it will also be more challenging to observe deviations
from $w=-1$ than if we assumed that $\delta w$ was constant and non-zero.  

Note that the initial values of $ $$\phi_{2}$ to which
the curves in Figure \ref{Hfitslowroll} correspond are roughly the same as the final values we have quoted in the Figure: ($\phi_{2,initial}\simeq\phi_{2,0})$.  This is because the Hubble friction ensures that $\phi_{2}$ does not roll down
much (see equation (15)). We
chose $\phi_{2,0}=16$ on the assumption that $\phi_{2,0}\simeq\phi_{2,initial}\simeq\phi_{1,initial}$
and $\phi_{1,initial}\simeq$16 because that is the minimum value
that can provide approximately $60$ e-folds of inflation. The other
values of $\phi_{2,0}$ were chosen to illustrate larger values of $\delta w_0$.

\begin{figure}
\includegraphics[width=84mm]{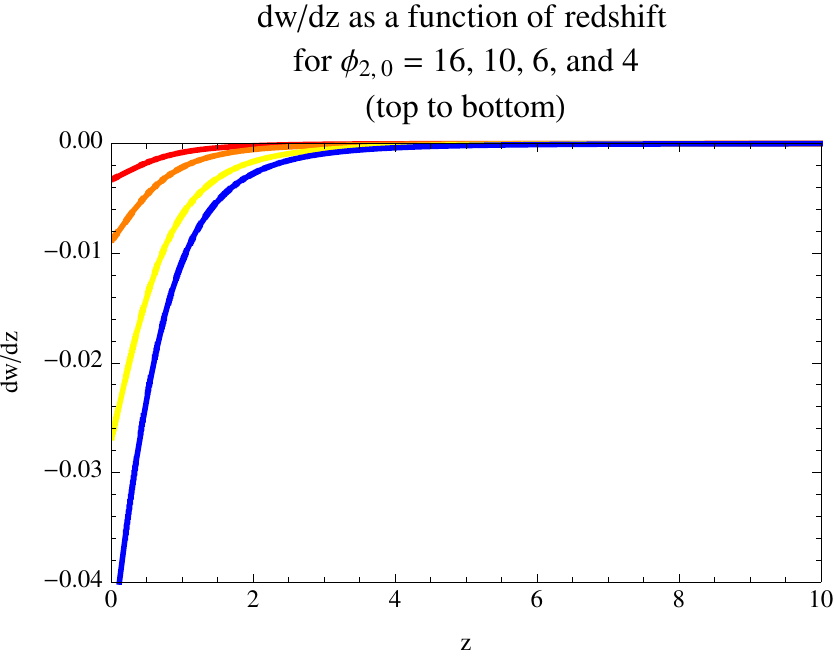}
\centering{}\caption{Plot of $dw/dz$ as a function of redshift. The values of $\phi_{2,0}$
are listed in the order that the curves go top to bottom ($\phi_{2,0}=16$
is the red curve, $\phi_{2,0}=10$ is the orange curve, etc.), and
chosen to be the same as those in Figure \ref{Hfitslowroll}. 
 \vspace{0.5in}}

\label{fig6}
\end{figure}

Figure \ref{fig6} shows $dw/dz$, which, since $w=\delta w-1$, is the same as $d\delta w/dz$.  This plot is consistent with what we might expect from Figure \ref{Hfitslowroll}: it is only for $z\lesssim 3$ that $dw/dz$ is significantly non-zero.

We note here that others who have studied a variety of quintessence
models near either maxima or minima in the potential have obtained
approximate analytical solutions for the evolution of $w(a)$ (Dutta \& Scherrer 2008, Dutta et al. 2009, Chiba et al. 2009, Huang, Bond, \& Kofman
2011). They have also concluded, as we will, that in these models
$w(a)$ is not well-fit by any linear function, including the popular
toy model $w(a)=w_{0}+w_{a}(1-a)$. Chiba in particular shows that
the results of Dutta and Scherrer should apply to a quadratic potential
such as the one we use.

\section{Conclusion}
In this paper, we have argued that the accelerated expansion in the early Universe (inflation) and the accelerated expansion today (dark energy) are caused by the same physical mechanism.  If inflation is slow-roll, then dark energy should also be slow-roll, and this allows us a chance to observe a detectable difference $\delta w$ between $w$ and $-1$.  We have suggested that the simplest model of inflation, Linde's chaotic inflation, where the potential is quadratic, could also effectively model dark energy.  For those who consider the super-Planckian values of the scalar field in this model problematic, we have suggested that the quadratic potential for inflation and the second quadratic potential we use for dark energy could both be realized as N-flation describing many sub-Planckian fields that are closely clustered in mass.  

The idea that dark energy and inflation are caused by the same mechanism produces a real gain in predictive power. Since we know the minimum value of the scalar field for inflation to give the 60 e-folds of inflation we see in the observable Universe, we can, on the reasonable assumption that the two scalar fields would have similar initial values following a Gaussian distribution, estimate upper limit probabilities for observing different values of $\delta w_0\equiv w_0+1$ today.

We also are able to use this slow-roll dark energy idea to make another prediction: that $\delta w\propto 1/H^2$.  Figure \ref{Hfitslowroll} shows, our numerical results seem to follow this formula. Our analytical work shows that this is generically true in any slow-roll model of dark energy.  This relationship can, we hope, provide a valuable template to observers seeking to look for a signal in noisy data.   Indeed, if observers find this shape for $w(z)$, it is a signature of slow-roll dark energy, which would be persuasive evidence that dark energy may be driven by a mechanism similar to that in inflation.  It may also motivate a focus on lower redshift observations, because this relationship implies that $\delta w$ will be close to zero for $z\gtrsim 3$.  

Finally, a third result we derive is an approximate formula for the value of $\delta w$ today as a function of the value of the dark energy scalar field today, $\phi_{2,0}$ (equation (\ref{boe})).  This formula is useful because if a deviation from $w=-1$ is measured, it can be plugged in to tell us what the value of $\phi_{2,0}$ is today in our model.  This, in turn, may tell us something about the initial value of the inflationary scalar field, as these fields are likely to have had initial values on the same order.  Having an idea of the inflationary scalar field's initial value may be interesting because it can tell us how many e-folds more than the minimum 60 we observe the Universe as a whole underwent during inflation.

Our model provides  a simple physical picture of what is causing dark energy, and complements it with a simple approximate formula for $\delta w\equiv w+1$.  We hope it will be useful as observers seek to interpret the upcoming data from BOSS (Baryon Oscillation Spectroscopic Survey) and $Planck$ as well as other future dark energy missions.

\section{Acknowledgements}

The authors thank Bharat Ratra and Gregory Novak for helpful conversations.  We also especially thank Paul Steinhardt for several fruitful discussions.  Finally, we are very grateful
to the anonymous referee for suggestions that improved the
scientific content and presentation of the paper.

\section{References}

\hspace{0.2in} Abramo L., Finelli F., 2003, Phys. Lett. B 575, 165

Albrecht A. et al.,  2006, Report of the
Dark Energy Task Force, arXiv:astro-ph/0609591

Armendariz-Picon C.,  Damour T., Mukhanov V., 1999, Phys. Lett. B 458, 209 

Armendariz-Picon C., Mukhanov V., Steinhardt P., 2000, Phys. Rev. Lett. 85, 4438 

Armendariz-Picon C., Mukhanov V.,  Steinhardt P., 2001, Phys. Rev. D 63, 103510 

Bagla J., Jassal H., Padmanabhan T., 2003, Phys. Rev. D 67, 063504 

Blandford R. et al., 2010, New Worlds, New Horizons in Astronomy and Astrophysics, National Academies Press

Cardenas V., 2006, Phys. Rev. D 73, 103512

Chevallier M., Polarski D., 2001, Int. J. Mod. Phys. D 10, 213 

Chiba T., 2009, Phys. Rev. D79, 083517

Cimatti A. et al., 2009, http://sci.esa.int/science-e/www/object/index.cfm?fobjectid=42822

Cline J.,  Jeon S., Moore G., 2004, Phys. Rev. D 70, 043543 

Copeland E., Sami M., Tsujikawa S., 2006, Int. J. Mod. Phys. D 15:1753-1936

de Simone A., Guth A. H., Linde A., Noorbala M., Salem M. P.,
Vilenkin A.,  2010, Phys. Rev.
D, 82, 6, 063520

Dicke R., Peebles P., 1979, in Hawking S., Israel W., eds, General Relativity: An Einstein Centenary Survey.  Cambridge University Press, Cambridge.

Dimopoulos S., Kachru S., McGreevy J., Wacker J.G., 2008,
JCAP 0808: 003

Dutta S., Scherrer R., 2008, Phys. Rev. D, 78, 123525

Dutta S., Saridakis E., Scherrer R., 2009, Phys. Rev. D, 79, 103005

Easther R., McAllister L., 2006, JCAP 0605: 018

Gasperini M., Piazza F., Veneziano G., 2002, Phys. Rev. D 65, 023508 

Gott J. R., 2008, Boltzmann Brains - I'd
rather see than be one, arXiv: 0802.0233

Guth A.H., 1981,
Physical Review D, 23, 2, 347-356 

Guth A.H., 2007, J. Phys. A40: 6811-6826

Guth A.H., Kaiser D., 2005, Sci. 307, 884Ð90

Huang Z., Bond J.R., Kofman L., 2011, ApJ 726, 64

Kallosh R., Linde A., 2010, JCAP 1011: 011

Kaloper N., Sorbo L., 2005, Phys. Rev. D, 74, 023513

Kamenshchik A., Moschella U.,  Pasquier V., 2001,
Phys. Lett. B 511, 265 

Kim S. A., Liddle R., 2006, Nflation: multifield
dynamics perturbations, arXiv: astro-ph/060560

Komatsu E., et al., 2010, Seven-Year Wilkinson Microwave
Anisotropy Probe (WMAP) Observations: Cosmological Interpretation,
Submitted ApJS

Li M., Li X., Zhang X., 2010, Sci. China Phys. Mech. Astron.53:1631-1645

Liddle A., Urena-Lopez L., 2006, Phys. Rev. Lett. 97, 161301

Linde A.D., 1983,
Phys. Lett. B. 129, 177 

Linde A. D., 2002, Inflationary Theory versus Ekpyrotic/Cyclic
Scenario, arXiv: hep-th/0205259

Linder E., 2003, Phys. Rev. Lett. 90, 091301 

Lyth D., Liddle A., 2000, Cosmological inflation and large-scale structure. Cambridge University Press, Cambridge.

Muller V., Schmidt H., 1989, General Relativity and Gravitation, Vol. 21 No. 5 

Padmanabhan T., 2002, Phys. Rev. D 66, 021301 

Park C., Kim Y.R., 2009, ApJL 715, 2

Peebles P., Vilenkin A., 1999, Phys.Rev. D 59, 063505 

Polarski D., Starobinsky A.A., 1992, Nuclear Physics B 385, 623-650 

Poletti S., 1989, Class. Quantum Grav. 6, 1943

Preskill J., 1979, Phys. Rev. Lett. 43, 1365Ð8 

Riess A. et al., 1998, AJ, 116: 1009-1038 

Riess A. et al., 2007, ApJ, 659:98-121

SDSS-III: Massive Spectroscopic Surveys of the Distant Universe, the Milky Way Galaxy, and Extra-Solar Planetary Systems, 2008, http://www.sdss3.org/science.php

Silk J., Turner M.S., 1987,
Physical Rev. D, Vol. 35, Issue 2, 419-428

Svrcek P., 2006, Cosmological Constant and Axions in
String Theory, arXiv:hep-th/0607086

Tegmark M. et al., 2004, Phys. Rev. D 69, 103501

Turner M., Villumsen J., Vittorio N., Silk J., Juszkiewicz R., 1987, ApJ, 323: 423-432 

Vilenkin A., 2003, International Journal of Theoretical Physics 42, 1193-1209

Weinberg S., 1987,
Phys. Rev. Lett. 59, 2607-2610 

Yamaguchi M., 2001, Phys. Rev. D 64, 063502 

Zunckel C., Gott J.R., Lunnan R., 2010, 412, 2, 1401-1408

\end{document}